\documentstyle[aps,twocolumn,epsf]{revtex}

\begin{document}
\draft
\title{Vortex phases in mesoscopic cylinders with \\
suppressed surface superconductivity}
\author{W.~V.~Pogosov}
\address{Moscow Institute of Physics and Technology, \\
141700 Dolgoprudnyi, Moscow region, Russia\\
}
\maketitle

\begin{abstract}
Vortex structures in mesoscopic cylinder placed in external magnetic field
are studied under the general de Gennes boundary condition for the order
parameter corresponding to the suppression of surface superconductivity. 
The Ginzburg-Landau equations are solved based on trial functions
for the order parameter for vortex-free, single-vortex, multivortex, and
giant vortex phases. The equilibrium vortex diagrams in the plane of external
field and cylinder radius and magnetization curves are calculated at 
different values of de Gennes "extrapolation length" characterizing the 
boundary condition for the order parameter. The comparison of the 
obtained variational results with some
available exact solutions shows good accuracy of our approach.
\end{abstract}


\tighten

\bigskip

\section{Introduction}

Recent achievements in electronic device miniaturization allow one to study
the mesoscopic superconducting samples with sizes of the order of the
coherence length $\xi (T)$. Such structures attract a considerable current
interest as a possible basis for low temperature electronics. The
superconducting state was studied experimentally for different-shaped
samples: discs, loops, double loops, dots etc. \cite{1,2,3,4}. It was shown
that the sample shape and sizes affect significantly the phase diagrams of
the mesoscopic superconductors.

The vortex phases in mesoscopic superconductors are commonly studied within
the framework of the Ginzburg-Landau theory \cite
{5,6,7,8,9,10,11,12,13,14,15,16,17}. As it is well-known from the
microscopic theory, the Ginzburg-Landau approach gives accurate results
provided that the order parameter undergoes only slight spatial variation on
the lenght scale of $\xi (0)$. This means that the Ginzburg-Landau theory
can be used in the temperature range not far from $T_{c}$. However, it is
known from the experience that it is also able to give reasonable 
results beyond this limit. The Ginzburg-Landau solutions for 
axially symmetric mesoscopic
samples (cylinders, discs) can be subdivided into two different types \cite
{6,7,8,9,10,11}. In the first case the modulus of the local order parameter
is axially symmetric inside the sample. The superconducting vortex-free
state, the single-vortex state, and the giant-vortex state belong to this
type of solutions. In the second case the axial symmetry is broken and a
vortex cluster is formed inside the sample (multivortex phase). This state
usually appears at lower fields and larger sample sizes as compared to the
giant-vortex phase \cite{7,8,10,11}. Note that multivortex state corresponds
to the Abrikosov flux-line lattice for the bulk superconductors.

The phase diagram of mesoscopic superconductor is strongly influenced by the
boundary condition for the order parameter. In general case it is given by
the de Gennes boundary condition \cite{18,19}: 
\begin{equation}
{\bf n}(-i{\bf \nabla }-{\bf A})\psi =\frac{i}{b}\psi \text{,}
\end{equation}
where ${\bf n}$ is the unit vector normal to the sample surface, $b$ is the
de Gennes ''extrapolation length'', ${\bf A}$ is the vector potential, and $%
\psi $ is the order parameter. Here and below the following dimensionless
variables are used: distances, magnetic field, and the order parameter are
measured in units of coherence length $\xi (T)$, bulk upper critical field $%
\allowbreak H_{c2}$, and $\sqrt{-\alpha /\beta }$, respectively, with $%
\alpha $, $\beta $ being the Ginzburg-Landau coefficients. The 
"extrapolation length" $b$ has a physical meaning of a length scale of 
the order parameter variation at the sample surface. Microscopic
considerations show that $b$ depends on the properties of interface, it is
maximum for an ideal surface with the mirror reflection of quasi-particles
and minimum for the rough surface with the diffusive reflection \cite
{18,19,20,21}. For the superconductor-dielectric (or vacuum) interface we
have $b\rightarrow \infty $\ in the former case and $b\sim \xi (0)$ in the
latter case. The value of $b$ depends also on the surface orientation 
provided that a superconducting gap is anisotropic. It follows from Eq. 
(1) that the order
parameter is suppressed in the vicinity of the sample surface at $b\sim \xi
(0)$. For the superconductor-normal metal interface $b$ is always small, $%
b\sim \xi (0)$, because of diffusion of normal electrons from the metal to
the superconductor. The ''extrapolation length'' $b$ in this case is a
function of temperature and various characteristics of metal and interface 
\cite{19,20,21,22,23}. There are also possibilities for the enhancement of
the order parameter at the interface that can be described by negative $b$
values. It can be realized by choosing the suitable material as a
surrounding medium \cite{24,25}, i.e., a superconductor having a higher
transition temperature then the material of the mesoscopic sample. Another
possibility is to use a semiconductor as a surrounding medium, such that
there is a overlap of the band gap of the semiconductor with the
superconducting gap. For the case of isotropic superconductor-vacuum ideal
interface ($b\rightarrow \infty $), the magnetic properties of mesoscopic
cylinders and discs, their equilibrium and non-equilibrium phase diagrams
were studied in numerous papers, see e.g. Refs. \cite
{5,6,7,8,9,10,11,12,13,14,15}, using different approaches. The vortex
structures allowing for the {\it enhanced} surface superconductivity were
studied in Refs. \cite{24,25}. In Ref. \cite{24} the case of mesoscopic
discs was considered within the lowest Landau level approximation, which
first was proposed in Refs. \cite{7,10}. In Ref. \cite{25} the
Ginzburg-Landau equations were solved numerically and self-consistently for
superconducting state in long cylinders.

In this paper we focus on magnetic properties of mesoscopic cylinders under
the general boundary condition for the order parameter corresponding to the
opposite case of {\it suppressed} surface superconductivity. For this
purpose, we propose a variational approach and solve the Ginzburg-Landau
equations without straightforward integration using trial functions for the
order parameter. These trial functions involve a set of variational
parameters yielding, as we show, more accurate quantitative description of
the spatial distribution of the order parameter than frequently used lowest
Landau level approximation. The approach is applicable to all vortex phases
(the vortex-free, the single-vortex, the multivortex, and the giant vortex
states) and any values of de Gennes ''extrapolation length''. The comparison
of variational calculations with some available exact results demonstrates
good accuracy of our approximation. The model enables us to calculate the
equilibrium $H_{e}$-$R$ diagram of the cylinder, where $H_{e}$ is the
external field and $R$ is the cylinder radius. The magnetization curves of
the cylinder are calculated.

\section{Model}

Let us consider a cylindrical type-II superconductor placed in the uniform
external magnetic field $H_{e}$ parallel to the cylinder axis. The sample is
assumed to be much longer than London penetration depth $\lambda (T)$.
Therefore, both the order parameter and the magnetic field are constant
along cylinder axis. We use the cylindrical coordinate system with
coordinates $r$, $\varphi $, $z$ and unit vectors ${\bf e}_{r}$, ${\bf e}%
_{\varphi }$, ${\bf e}_{z}$.

The system of Ginzburg-Landau equations is given by \cite{26}: 
\begin{equation}
|\psi |^{2}\psi -\psi +\left( i{\bf \nabla +A}\right) ^{2}\psi =0,
\end{equation}
\begin{equation}
\text{rot\ }{\bf H=}\frac{1}{\kappa ^{2}}\left[ {\bf A}|\psi |^{2}+\frac{i}{2%
}\left( \psi ^{\ast }{\bf \nabla }\psi -\psi ^{\ast }{\bf \nabla }\psi
\right) \right] ,
\end{equation}
where ${\bf H}$ and $\psi $ are the dimensionless local magnetic field and
the order parameter (${\bf H}=$ rot ${\bf A}$, ${\bf H}=H{\bf e}_{z}$); $%
\kappa =\lambda (T)/\xi (T)$ is the Ginzburg-Landau parameter. Equations (2)
and (3) must be supplemented by the boundary conditions for the order
parameter (1) and the magnetic field: 
\begin{equation}
H(R)=H_{e}.
\end{equation}
Next, we expand all variables in powers of $\kappa $:

\begin{equation}
\psi =\sum_{n=0}^{\infty }\psi _{2n}\frac{1}{\kappa ^{2n}}\text{, \ \ }{\bf A%
}=\sum_{n=0}^{\infty }{\bf A}_{2n}\frac{1}{\kappa ^{2n}}\text{, \ \ }%
H=\sum_{n=0}^{\infty }H_{2n}\frac{1}{\kappa ^{2n}}.
\end{equation}

We substitute expansions (5) to Ginzburg-Landau equations (2), (3) and to
boundary conditions (1), (4) and equate powers of $\kappa $. It is easy to
show that the vector potential and the magnetic field at leading order are
given by:

\begin{equation}
{\bf A}_{0}={\bf e}_{\phi }\frac{H_{e}r}{2}\text{, \ \ \ \ \ \ \ }H_{0}=H_{e}%
\text{.}
\end{equation}
The order parameter at leading order $\psi _{0}$ is determined by the first
Ginzburg-Landau equation (2) and the boundary condition (1) at ${\bf H}={\bf %
H}_{0}$. This condition is now given by:

\begin{equation}
\frac{\partial \psi _{0}}{\partial r}+\frac{\psi _{0}}{b}=0\text{.}
\end{equation}
In the next order one has from Eq. (3):

\begin{equation}
H_{2}=-\int_{0}^{r}dr{\bf e}_{\varphi }\left[ |\psi _{0}|^{2}{\bf A}_{0}+%
\frac{i}{2}\left( \psi _{0}^{\ast }{\bf \nabla }\psi _{0}-\psi _{0}^{\ast }%
{\bf \nabla }\psi _{0}\right) \right] \text{.}
\end{equation}
Thus, at leading order the magnetic field is uniform inside the sample and
the magnetization equals zero. Physically, this implies that the additional
field generated by the Meissner current and by the vortices is of the order
of $1/\kappa ^{2}$ in comparison with the external uniform field. Below, we
shall calculate the energy of the sample at leading order, whereas the
magnetization will be found at the next order using Eq. (8). It was shown in
Ref. \cite{27} that this approximation is accurate not only for high-$\kappa 
$ materials but also for moderate-$\kappa $ superconductors (e.g. $\kappa
\approx 4$) with sizes comparable to $\xi (T)$.

We present $\psi _{0}$ as a Fourier series: 
\begin{equation}
\psi _{0}(r,\varphi )=\sum_{L_{j}=0}^{\infty }f_{L_{j}}(r)\exp
(-iL_{j}\varphi )\text{.}
\end{equation}
For the axial symmetric distribution of the modulus of the order parameter
inside the sample the only one term in Eq. (9) is nonzero. The vortex-free,
single-vortex, and giant vortex phases with angular quantum momentum $L$
correspond to the harmonics with $L_{j}=0$, $1$, and $L$, respectively. The
modulus of the order parameter in the multivortex phase is not axially
symmetric. Note that in this case the symmetry of the vortex configuration
imposes a restriction on functions $f_{L_{j}}$: some of these functions
equal zero. It was shown in Refs. \cite{7,10} that taking into account only
two main terms in right-hand side of Eq. (9) is enough for an accurate
calculation of the energy of the mesoscopic superconductor in multivortex
state. The vortex cluster with $L$ vortices on a ring and no vortex at the
axis (ring-like configuration) can be described as a mixture of two
components with $L_{1}=0$ and $L_{2}=L$ (($0:L$) state). The vortex cluster
with one vortex at the cylinder axis and $(L-1)$ vortices on a ring
corresponds to $L_{1}=1$ and $L_{2}=L$ (($1:L$) state). The contribution
from other harmonics is small and can be neglected, when we consider
few-fluxoid cylinders that can accommodate just few vortices \cite{7,10}.

Using Eqs. (2) and (9) it can be easily shown that each function $%
f_{L_{j}}(r)$ has the following asymptotic at $r\rightarrow 0$:

\begin{equation}
f_{L_{j}}(r)\sim r^{L_{j}}\text{.}
\end{equation}
Besides, each function $f_{L_{j}}(r)$ meets boundary condition (7). These
conditions for $f_{L_{j}}(r)$ are valid both for the giant vortex and the
multivortex phases. It is a rather complicated task to find $f_{L_{j}}(r)$
from the first Ginzburg-Landau equation (2) due to its non-linearity.
Instead of the straightforward integration of this 
equation, it is possible to use trial functions for the coordinate
dependence $f_{L_{j}}(r)$. Notice that different variational procedures
allowing one to solve approximately the Ginzburg-Landau equations were used
in numerous papers for mesoscopic \cite{7,10,15}, bulk \cite{28,29,30,31},
and different-shaped \cite{32} superconductors. One can easily show that if 
Eq. (9) is a solution of the first Ginzburg-Landau equation, each
function $f_{L_{j}}(r)$ can be represented as the following series expansion
in powers of $r/R$\ without loss of generality:

\begin{equation}
f_{L_{j}}(r)=\exp \left( -q_{L_{j}}\frac{r^{2}}{R^{2}}\right)
\sum_{u=0}^{\infty }p_{u}^{L_{j}}\left( \frac{r}{R}\right) ^{L_{j}+2u},
\end{equation}
where $p_{u}^{L_{j}}$\ are constants. The value of $q_{L_{j}}$ can be found 
from boundary condition (7):

\begin{equation}
q_{L_{j}}=\frac{R}{2b}+\frac{1}{2}\frac{\sum_{u=0}^{\infty
}p_{u}^{L_{j}}(L_{j}+2u)}{\sum_{u=0}^{\infty }p_{u}^{L_{j}}}\text{.}
\end{equation}
Our approach is to consider the coefficients $p_{u}^{L_{j}}$ as variational
parameters and to minimize the free energy with respect to $p_{u}^{L_{j}}$.
The exponential prefactor in Eq. (11) takes into account the suppression 
of the order
parameter at the contact with a surrounding material. First term in the 
expansion describes the behavior of the order 
parameter in the central part of the sample. Other terms specify 
the order parameter in the sample as a whole.
As it will be shown below, taking into account only first three terms in Eq.
(11) is enough for the accurate calculation of the order parameter
distribution, and we will use this approximation in all calculations of
magnetization and $H_{e}$-$R$ diagrams. 
Note that trial function (11) was used in Ref. \cite{32} for
the study of surface superconductivity in samples of different complex
shapes placed in vacuum ($b\rightarrow \infty $).

The Ginzburg-Landau functional for the Gibbs free energy $G$ of the cylinder
can be written as a sum of two contributions, $G_{b}$ and $G_{s}$. The
former is the bulk energy of the sample and the latter is the surface
energy. These contributions are given by \cite{20,23,26}: 
\begin{eqnarray}
G_{b} &=&\int \Biggl[-|\psi |^{2}+\frac{1}{2}|\psi |^{4}+|\left( -i{\bf %
\nabla }-{\bf A}\right) \psi |^{2}+  \nonumber \\
&&\kappa ^{2}{\bf H}^{2}-2\kappa ^{2}{\bf HH}_{e}\Biggr]dV\text{,}
\end{eqnarray}

\begin{equation}
G_{s}=\frac{1}{b}\int |\psi |^{2}dS\text{.}
\end{equation}
The integration in Eqs. (13) and (14) is performed over the sample bulk and
surface, respectively. Note that the general boundary condition for the
order parameter (1) can be obtained phenomenologically by minimization of
the free energy functional $G=G_{b}+G_{s}$ with respect to the order
parameter $\psi $ and the vector potential ${\bf A}$ \cite{20,23,33}.

Substituting expansion (9) to Eqs. (13) and (14) and taking into account Eq.
(6) we obtain the expression for the energy of the multivortex state (per
unit length of the cylinder): 
\begin{eqnarray}
G_{b} &=&2\pi \int_{0}^{R}rdr\Biggl[\frac{1}{2}\left(
f_{L_{1}}^{4}+f_{L_{2}}^{4}+4f_{L_{1}}^{2}f_{L_{2}}^{2}\right) -  \nonumber
\\
&&f_{L_{1}}^{2}-f_{L_{1}}^{2}+\left( \frac{df_{L_{1}}}{dr}\right)
^{2}+\left( \frac{df_{L_{2}}}{dr}\right) ^{2}-\kappa ^{2}H_{e}^{2}+ 
\nonumber \\
&&f_{L_{1}}^{2}\left( \frac{H_{e}r}{2}-\frac{L_{1}}{r}\right)
^{2}+f_{L_{2}}^{2}\left( \frac{H_{e}r}{2}-\frac{L_{2}}{r}\right) ^{2}\Biggr],
\end{eqnarray}
\begin{equation}
G_{s}=\frac{2\pi R}{b}\left( f_{L_{1}}^{2}(R)+f_{L_{2}}^{2}(R)\right) .
\end{equation}
The energy of any state having axially symmetric modulus of the order
parameter with angular momentum $L$ reduces to Eqs. (15) and (16). In this
case we must put $L_{1}=L,$ $f_{L_{2}}=0$. Using Eqs. (11) and (12) we find
the energy $G$ from Eqs. (15) and (16) by a straightforward integration as
an explicit function of variational parameters $p_{0}^{L_{1}}$, $%
p_{1}^{L_{1}}$, $p_{2}^{L_{1}}$\ and $p_{0}^{L_{2}}$, $p_{1}^{L_{2}}$, $%
p_{2}^{L_{2}}$. The resulting expression, however, is rather cumbersome and
we do not present it here. Finally, values of the variational parameters at
each $R$ are found numerically by the minimization of the free energy. This
procedure yields the local order parameter and the energy of the cylinder.

Knowing the local order parameter we can calculate the magnetization. It is
given by:

\begin{equation}
-4\pi M=<H>-H_{e}\text{,}
\end{equation}
where $<H>$ is the averaged magnetic field over the superconductor volume.
Taking into account Eq. (8) and expansion (9) we obtain:

\begin{eqnarray}
-4\pi M &=&\frac{2}{\kappa ^{2}R^{2}}\int_{0}^{R}rdr\int_{0}^{r}dx\Biggl[%
f_{L_{1}}^{2}(x)\left( \frac{H_{e}x}{2}-\frac{L_{1}}{x}\right) +  \nonumber
\\
&&f_{L_{2}}^{2}(x)\left( \frac{H_{e}x}{2}-\frac{L_{2}}{x}\right) \Biggr]%
\text{.}
\end{eqnarray}
Functions $f_{L_{1}}$\ and $f_{L_{2}}$\ are found by the method described
above. Hence, one can calculate the magnetization using Eq. (18). In the
following section we apply the developed approach for the analysis of the
behavior of the cylinder in the external field.

\section{Results and discussion}

Comparing the energies of different states one can calculate the equilibrium 
$H_{e}$-$R$ diagram of the cylinder. The results of our calculations are
shown in Fig. 1 for different $b$ values: $b=1$ (a), $b=2.5$ (b), $b=5$ (c),
and $b\rightarrow \infty $ (d). The latter case corresponds to the isotropic
superconductor-vacuum ideal interface, and was studied in Refs. \cite
{5,6,10,11,13,14,15}. Curve 1 shows the transition from the normal to the
superconducting state (the surface critical field $H_{c3}$). The oscillatory
behavior of the function $H_{c3}(R)$ is caused by the fact that the
transition occurs from the normal to the giant vortex states with different
angular quantum moments $L$ depending on the cylinder radius. Besides, the
function $H_{c3}(R)$ depends appreciably on the value of $b$: the value of $%
H_{c3}(R)$ decreases with decrease of $b.$ At $R\rightarrow \infty $ the
dependence $H_{c3}(R)$ tends to the surface critical field for the
half-space sample, which was calculated in Ref. \cite{22} as a function of $%
b $. Note that $H_{c3}$ $(\infty )=1.695$ for $b\rightarrow \infty $ and $%
H_{c3}(\infty )$ $=1$ for $b\rightarrow 0$ \cite{18,22}.

Below $H_{c3}$ the transitions between different giant vortex states take
place. Solid lines show the phase boundaries between the states with
different vorticities that are the sums of angular quantum moments of all
vortices and giant vortices. In the giant vortex state the order parameter
is strongly suppressed in the inner part of the cylinder, and this state can
be referred to a surface superconductivity. For illustration, the spatial
distribution of the order parameter in the giant vortex phase with $L=2$ is
plotted in Fig. 2 at $b=1$, $R=3.9$, $H_{e}=0.9$ (a) and $b\rightarrow
\infty $, $R=4.56$, $H_{e}=0.5$ (b) (solid lines). In the former case the
order parameter is also suppressed at the sample surface because of
the small $b$
value (e.g., superconductor-normal metal interface). To check the accuracy
of our approach we took into account next several terms in expansion (11),
thus increasing the number of variational parameters. As we found, this
practically did not change the calculated order parameter for almost all
points of $H_{e}$-$R$ diagram shown in Fig. 1. This result implies that our
variational calculations are close to the exact solutions of the
Ginzburg-Landau equations since expansion (11) is written without any loss
of generality. Therefore, one can find Ginzburg-Landau solution with any
desired accuracy (for the states with axially symmetric modulus 
of the order parameter) allowing for enough number of variational parameters $%
p_{u}^{L_{j}}$. For multivortex states the accuracy is limited by the 
fact that we take into account only two main harmonics in the Fourier 
expansion (9). In Fig. 2 we also plotted the spatial variation of the order 
parameter calculated within the lowest Landau level approximation (dashed
lines). In this approximation the order parameter is proportional to the
eigenfunction of the kinetic energy operator corresponding to the lowest
eigenvalue. We found that this approach remains very accurate not far from $%
H_{c2}$ and $H_{c3}$ (see Fig. 2(a)). At lower fields the results of the
lowest Landau level approximation are not so accurate (Fig. 2(b)).

As follows from Fig. 1, the superconducting state does not nucleate at very
small cylinder radii, smaller than some critical radius, and the sample
is in normal state at any applied field. The critical radius tends to zero
at $b\rightarrow \infty $. There is also the interval of $R$ for each $b$,
when the vortex phase does not nucleate, and the transition occurs from the
normal to the superconducting vortex-free state. Every vortex phase with
vorticity $L>1$ can exist in the form of the giant vortex or the multivortex
configuration. The dashed lines on Fig. 1 show the boundaries between these
states. Below these lines for given $L>1$ the multivortex state has the
lowest energy, and above these curves the giant vortex state becomes more
energetically favorable. In equilibrium state the multivortex phase can
exist if the applied field is smaller than $1$ ($H_{c2}$ in dimensional
units) and if the radius of the cylinder is large enough. For each $b$ there
exists an interval of small cylinder radii, when the multivortex phase is
energetically unfavorable as compared to the states with axial symmetric
distribution of the modulus of the order parameter. Curve 2 in Fig. 1 shows
the lower critical field that corresponds to the equilibrium boundary
between the vortex-free and the single-vortex states.

With increasing the external field the cylinder can follow rather complex
set of phase transitions. It can come from the giant vortex to the
multivortex states and then back to the giant vortex phase. For example, at $%
R=4.4$, $b=1$ the following set of transitions occurs: $0\rightarrow
1\rightarrow (0$:$2)\rightarrow 2\rightarrow (0$:$3)\rightarrow 3\rightarrow
(0$:$4)\rightarrow 4$. At $R=4.1$, $b=\infty $ the set of transitions is: $%
0\rightarrow 1\rightarrow 2\rightarrow (0$:$3)\rightarrow 3\rightarrow (0$:$%
4)\rightarrow 4\rightarrow 5\rightarrow 6\rightarrow 7\rightarrow
8\rightarrow 9\rightarrow 10\rightarrow 11$.

The transitions between the phases with different vorticities are always
discontinuous, they occur when the energies of different states become
equal. The transitions between the states with the same vorticity may be
continuous as well as discontinuous. Continuous phase transitions occur
between the multivortex and the giant vortex states. In this case, with
increasing of the applied field the intervortex distances decrease, and
vortices merge into the giant vortex located at the cylinder axis. The
similar results were obtained in Refs. \cite{7,8,9,10,11}\ for the case 
$b=\infty $ and in Ref. \cite{24}\ at $b<0$\ (enhanced surface
superconductivity). In our calculations we also took into account a
possibility of existence of the multivortex clusters with central vortex. It
turned out that such configurations can be more favorable energetically than
giant vortex states in some regions of $H_{e}$-$R$ diagram shown in Fig. 1.
However, in all these cases the ring-like clusters have the lower energy.
For cylinders thicker than shown in Fig. 1 we found that at any $b>0$ the
ground state can be represented by configurations with the central vortex.
The transitions between multivortex states of the same vorticity with and
without central vortex are discontinuous.

With increasing of the cylinder radius the number of vortices, which it is
able to accomodate, increases, and finally vortex array transforms to
the classical triangular Abrikosov flux-line lattice far from the surface.
The first step on this
way is the appearance of the clusters with central vortex. However, we do
not analyze here the transition from the mesoscopic to the macroscopic
behavior and restrict ourselves on few-fluxoid cylinders, which can
accomodate only few vortices before the thansition to the normal state and
whose magnetic properties can be described in terms of mixture of only two
harmonics in Eq. (9).

Now we find the magnetization of the cylinder using Eq. (18). The results of
our calculations at $\kappa =5$ are presented in Fig. 3(a) for $b=1$, $%
R=4.625$ and in Fig. 4(a) for $b\rightarrow \infty $, $R=4.05$. In these
cases the cylinders accommodates the giant vortices with maximum angular
quantum moments equal to $5$ and $11$, respectively, before the transition
to the normal state. Jumps in the magnetization correspond to the
transitions between the states with different vorticity. It is interesting
that in the state with suppressed surface superconductivity (Fig. 3(a)) the
discontinuity of the field dependence of the magnetization is less
pronounced, especially at high fields. In the first case (Fig. 4(a))
at $H_{e}\approx 0.61
$ ($0.61H_{c2}$ in dimensional units) the transition occurs from the
multivortex state with $2$ vortices to the giant vortex state with angular
quantum momentum $2$ ($2\rightarrow (0$:$2))$. In the second case (Fig. 4(b))
the transition $3\rightarrow (0$:$3)$ occurs at $H_{e}\approx 0.74$ and the
transition $4\rightarrow (0$:$4)$ occurs at $H_{e}\approx 0.87$. All these
transitions are followed by weak jumps in the slope of the magnetization.
The behavior of the magnetization near the transitions is shown in Fig. 3(b)
and 4(b). Solid lines denote the equilibrium magnetization, dashed lines
denote the metastable magnetization corresponding to the giant vortex phase.
The similar behavior of the magnetization in the vicinity of the
multivortex-giant vortex transitions was reported in Ref. [34] for the case $%
b\rightarrow \infty $.

Next we discuss the accuracy of our variational procedure. First, we compare
the variational results with known exact solutions for the surface critical
field. The phases with axial symmetric distribution of the modulus of the
order parameter are always energetically more favorable with respect to the
multivortex state at applied field higher than the bulk upper critical field
(see phase diagrams on Fig. 1). In the vicinity of the surface critical
field the first Ginzburg-Landau equation (1) can be linearized. A resulting
equation has the following analytical solution \cite{5}:

\begin{equation}
f_{L}(R)=r^{L}\exp \left( -\frac{H_{e}r^{2}}{4}\right) \Phi \left( \frac{%
H_{e}-1}{2H_{e}},L+1,\frac{H_{e}r^{2}}{2}\right) \text{,}
\end{equation}
where $\Phi $ is Kummer function. Function (19) must meet the boundary
condition (7). This yields the transcendental equation for the surface
critical field allowing one to find $H_{c3}(R)$ exactly. The comparison of
the variationally calculated $\frac{{}}{{}}H_{c3}$ with this exact
dependence shows good agreement with an accuracy better than $1\%$ for all
values of $b$ and $R$ under study. The lower critical field of the cylinder $%
H_{c1}$ versus $R$ was calculated numerically in Ref. \cite{12} at $%
b\rightarrow \infty $. The comparison of this result with our dependence $%
H_{c1}(R)$ reveals the same accuracy. Thus, our results appear to be a good
approximation to the exact Ginzburg-Landau solutions for the mesoscopic
cylinders.

In summary, we analyzed the superconducting state in long mesoscopic
cylinder with suppressed surface superconductivity. An asymptotic expansion
was used to simplify the Ginzburg-Landau equations at high and moderate
values of $\kappa $, and the simplified equations were solved by variational
method. The equilibrium $H_{e}$-$R$ diagram of the cylinder were obtained,
where $H_{e}$ is the external field and $R$ is the cylinder radius, at
different values of ''extrapolation length''. We showed that magnetic
properties of the cylinder depend appreciably on the value of
''extrapolation length'' and studied the evolution of $H_{e}$-$R$ diagram
with changing of ''extrapolation length''.

\section*{ACKNOWLEDGMENTS}

The author acknowledges useful discussions with K. I. Kugel, A. L.
Rakhmanov, and E. A. Shapoval. This work was supported by the Russian
Foundation for Basic Research (RFBR), grants Nos. 00-02-18032 and
01-02-06526, by the joint INTAS-RFBR program, grant No. IR-97-1394, and by
the Russian State Program 'Fundamental Problems in Condensed Matter Physics'.

\onecolumn
\newpage

\section*{FIGURE CAPTIONS}

\bigskip

Fig.1. Equilibrium $H_{e}$-$R$ diagram of the cylinder in the external
magnetic field at $b=1$ (a), $b=2.5$ (b), $b=5$ (c), $b\rightarrow \infty $
(d). Solid lines show the boundaries between the states with different
vorticity. Dashed lines correspond to the boundaries between the multivortex
and the giant vortex phases. Curves 1 and 2 show the surface and the first
critical fields, respectively. The dot line denotes the bulk upper critical
field. \bigskip

Fig 2. The spatial distribution of the dimensionless modulus of the order
parameter inside the mesoscopic cylinder in giant vortex state with angular
quantum momentum $L=2$ at $b=1$, $R=3.9$, $H_{e}=0.9$ (a) and $b\rightarrow
\infty $, $R=4.56$, $H_{e}=0.5$ (b) \ Solid lines correspond to our
variational result, dashed lines to the results of the lowest Landau level
approximation. The distance from the cylinder axis $r$ is measured in units
of the coherence length $\xi (T)$. \bigskip 

Fig. 3. The equilibrium magnetization of the cylinder with radius $R=4.625$
versus applied field at $b=1$, $\kappa =5$. Jumps in the magnetization in
Fig. 3(a) correspond to the transitions between the states with different
vorticity. Fig. 3(b) shows the behavior of the magnetization in the vicinity
of the continuous phase transition at $H_{e}\approx 0.61H_{c2}$ from the
multivortex state with $2$ vortices to the giant vortex phase with vorticity 
$L=2$. Solid lines correspond to the equilibrium magnetization, dashed line
shows the magnetization of the metastable giant vortex phase. \bigskip

Fig. 4. The equilibrium field dependence of the magnetization of the
cylinder with radius $4.05$ at $b\rightarrow \infty $, $\kappa=5$. In Fig.
4(b) the magnetization is plotted versus applied field near the transitions
from the multivortex states to the giant vortex phases. Solid lines
correspond to the equilibrium magnetization, dashed line shows the
magnetization of the metastable giant vortex phase.

\newpage

\begin{figure}[tbp]
\end{figure}

\newpage

\begin{figure}[tbp]
\end{figure}
\newpage

\begin{figure}[tbp]
\end{figure}
\newpage

\begin{figure}[tbp]
\end{figure}

\newpage
\begin{figure}[tbp]
\epsfxsize= .85\hsize
\centerline{ \epsffile{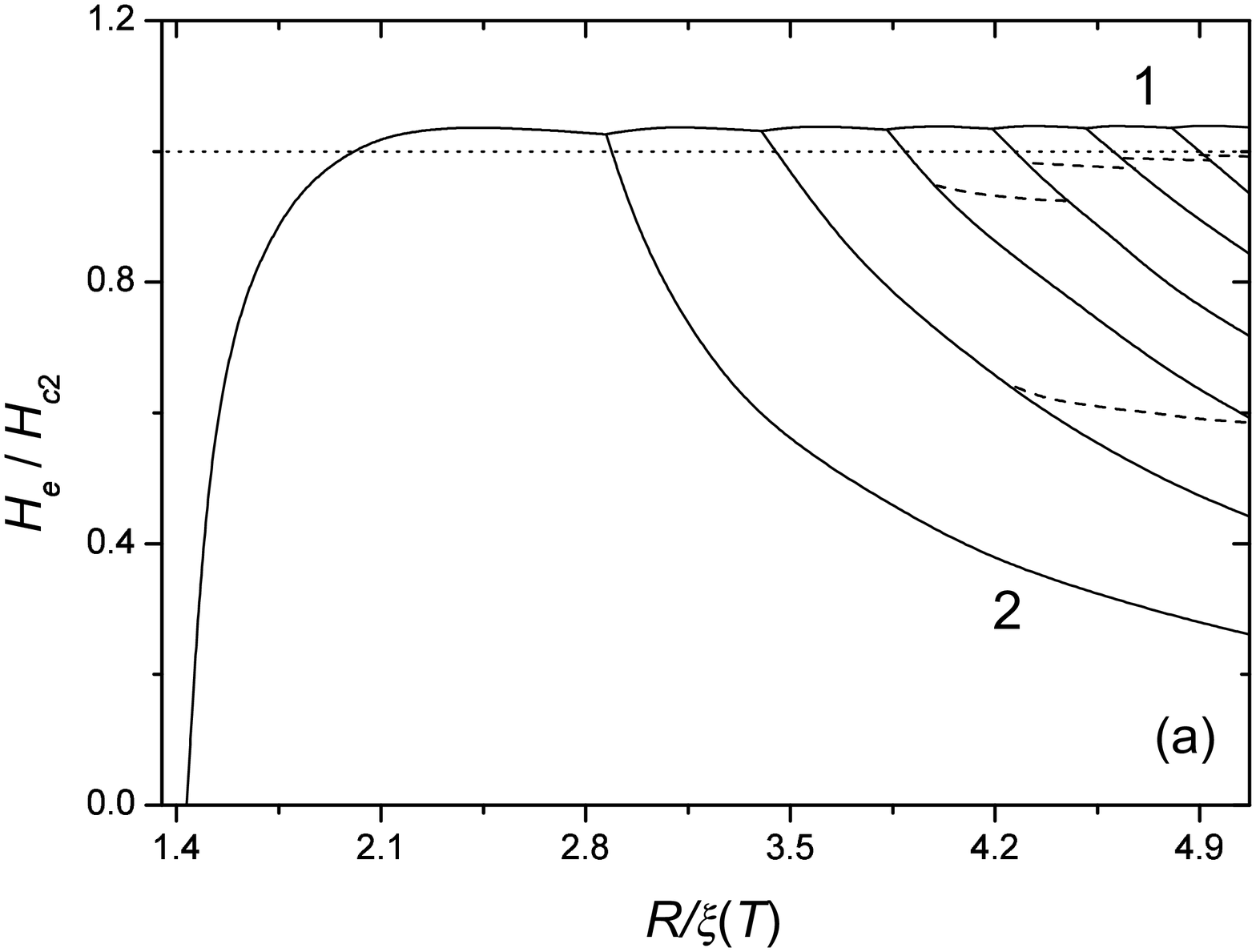}}
\text{Fig. 1(a)}
\epsfxsize= .85\hsize
\centerline{ \epsffile{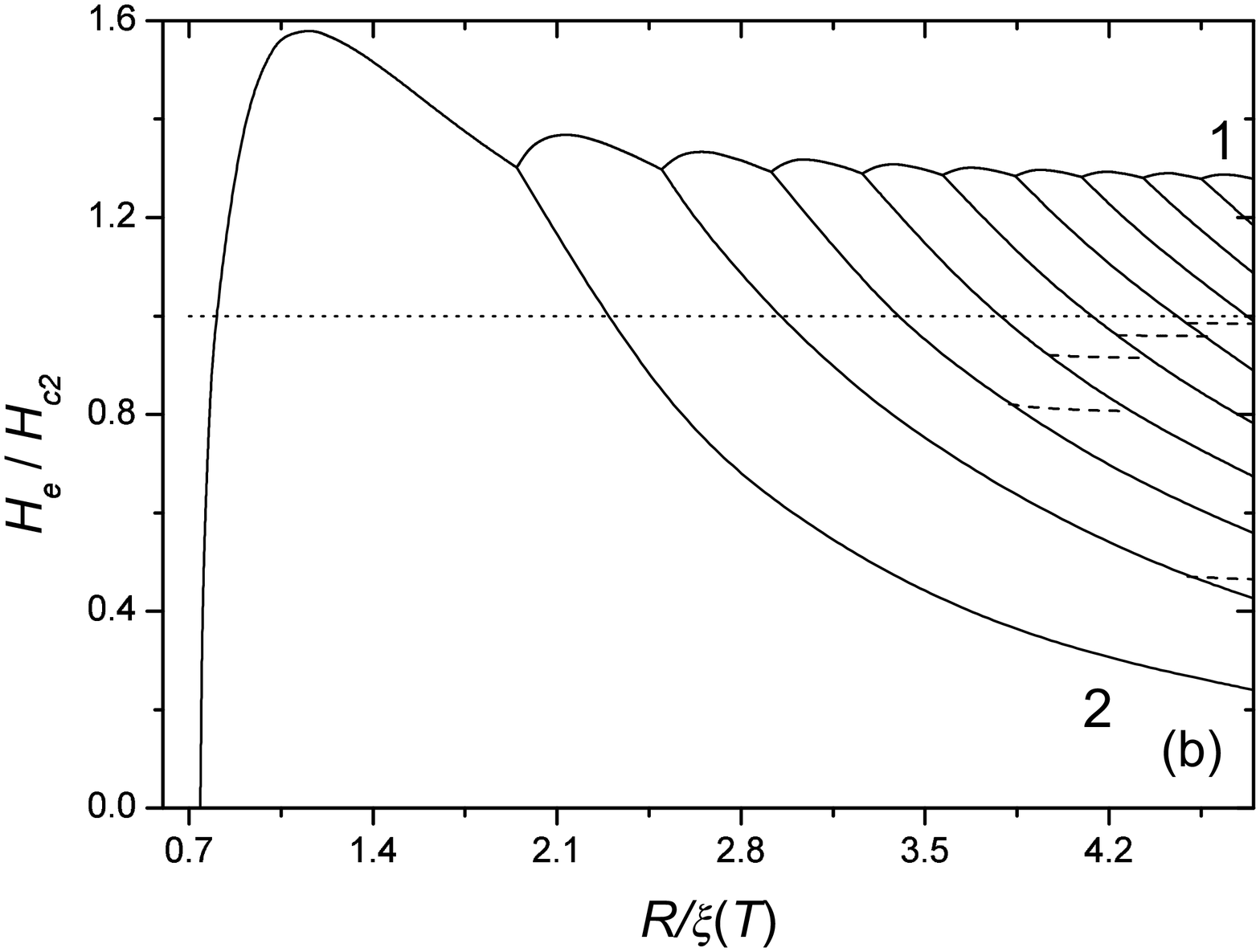}}
\text{Fig. 1(b)}
\newpage
\epsfxsize= .85\hsize
\centerline{ \epsffile{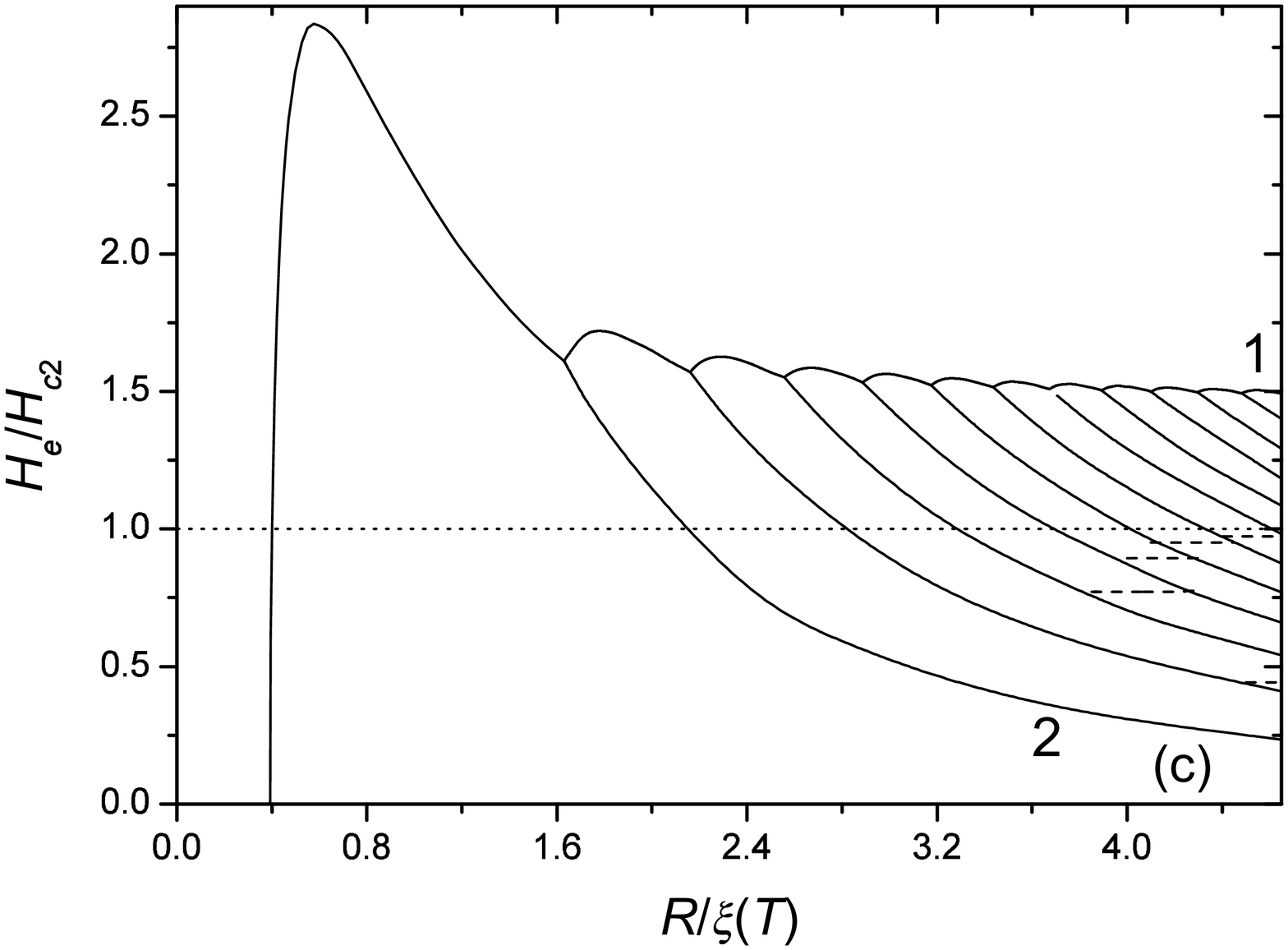}}
\text{Fig. 1(c)}
\epsfxsize= .85\hsize
\centerline{ \epsffile{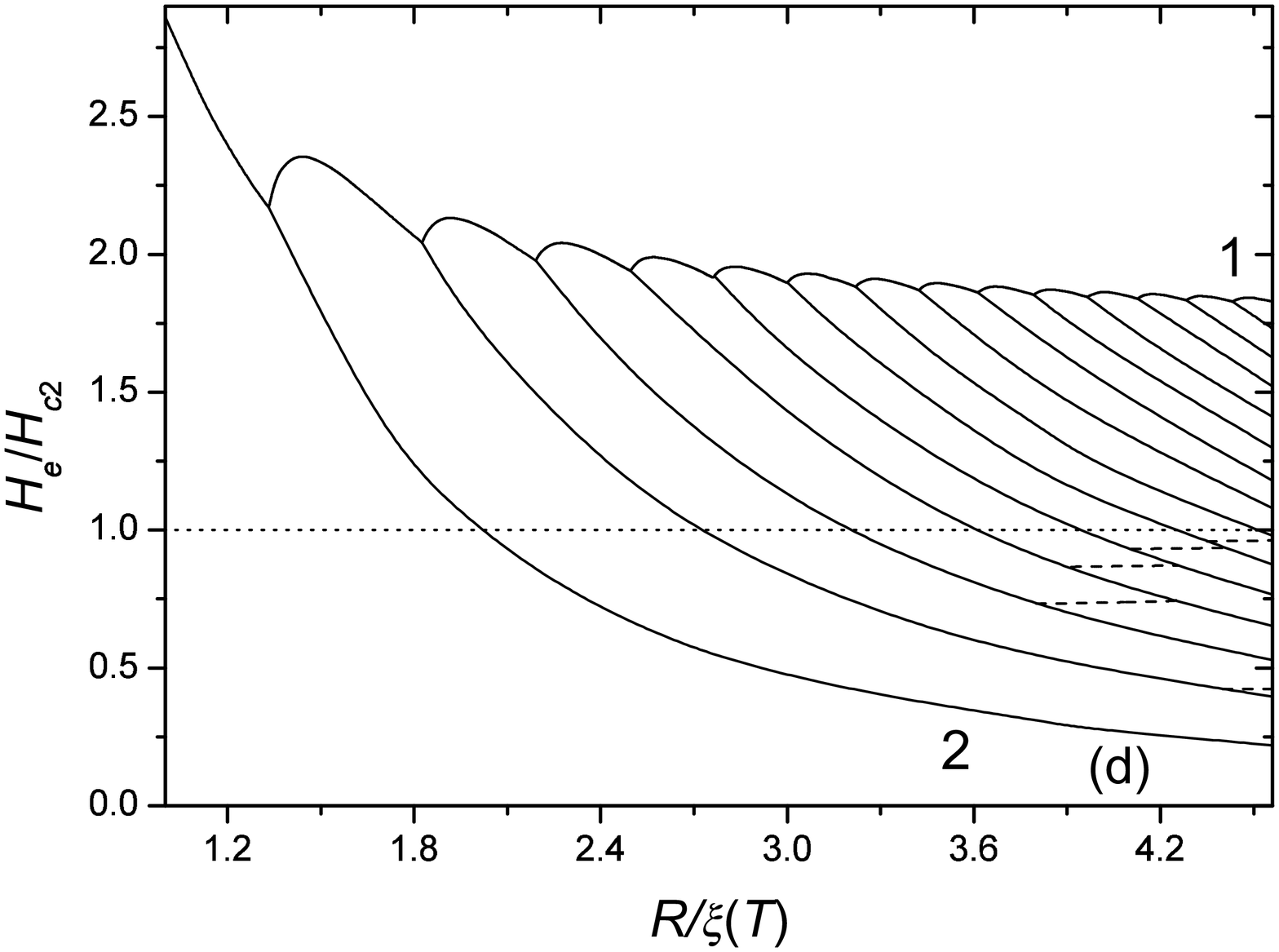}}
\text{Fig. 1(d)}
\end{figure}

\newpage
\begin{figure}[tbp]
\epsfxsize= .8\hsize
\centerline{ \epsffile{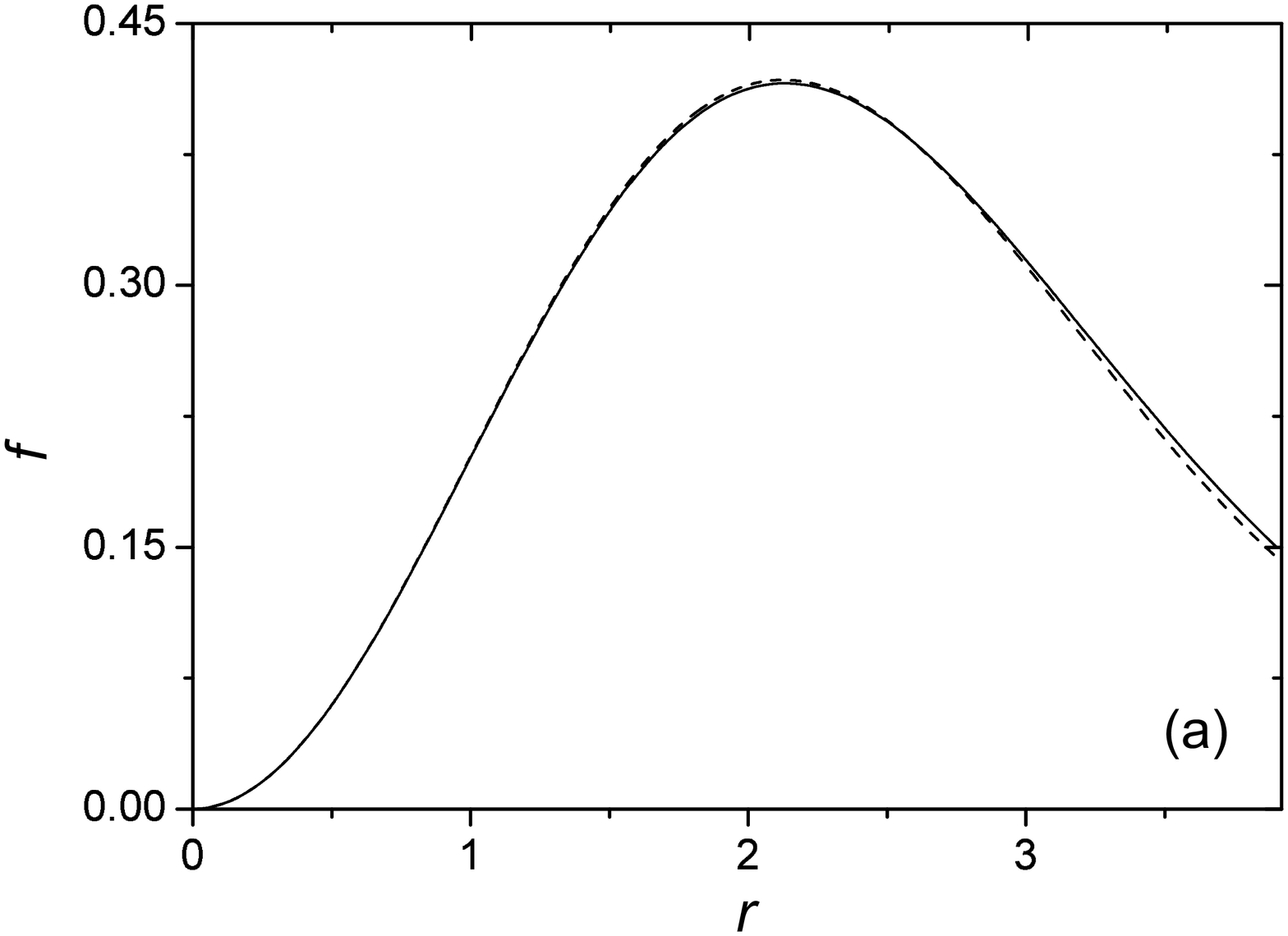}}
\text{Fig. 2(a)}
\epsfxsize= .8\hsize
\centerline{ \epsffile{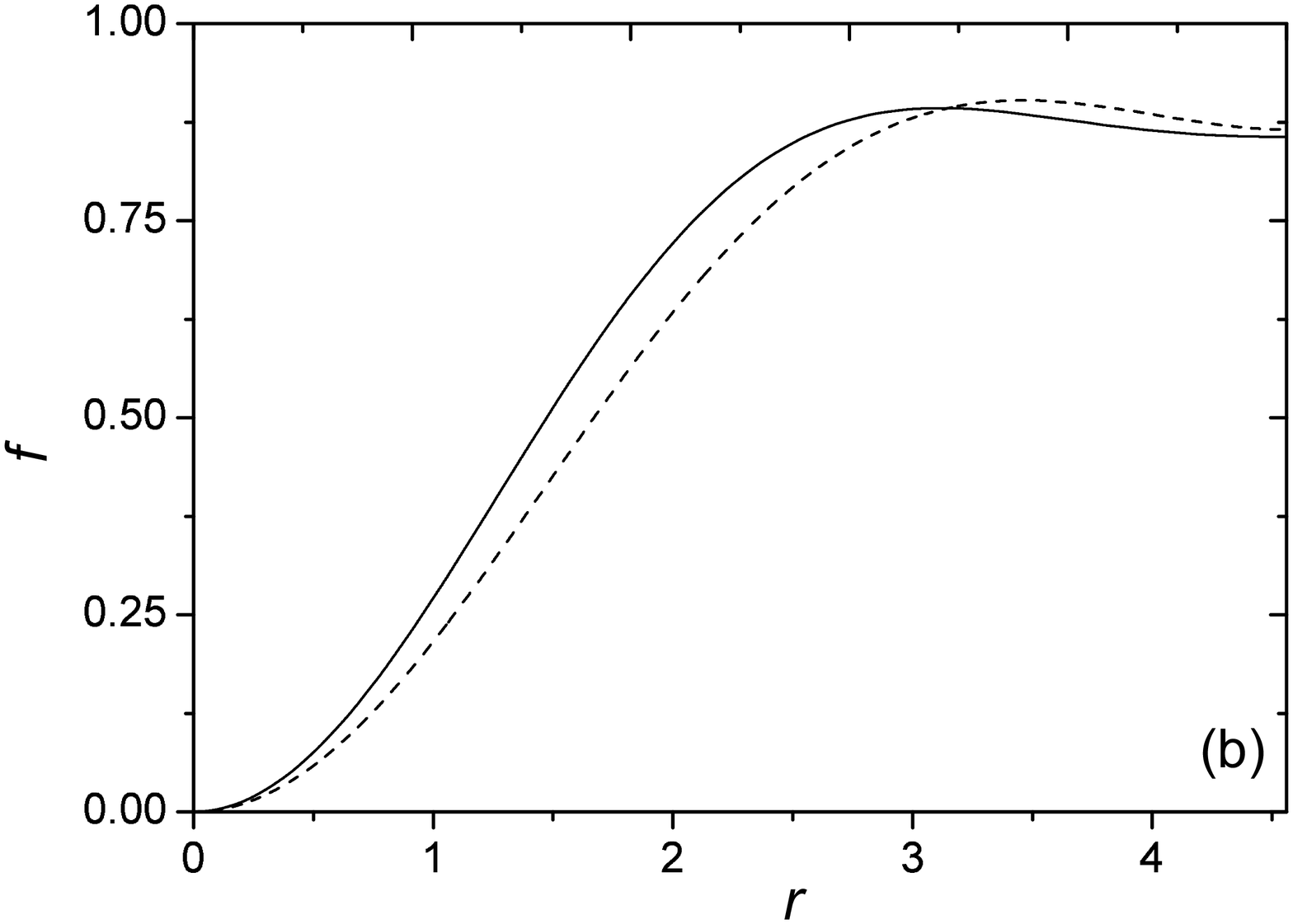}}
\text{Fig. 2(b)}
\end{figure}
\newpage
\begin{figure}[tbp]
\epsfxsize= .85\hsize
\centerline{ \epsffile{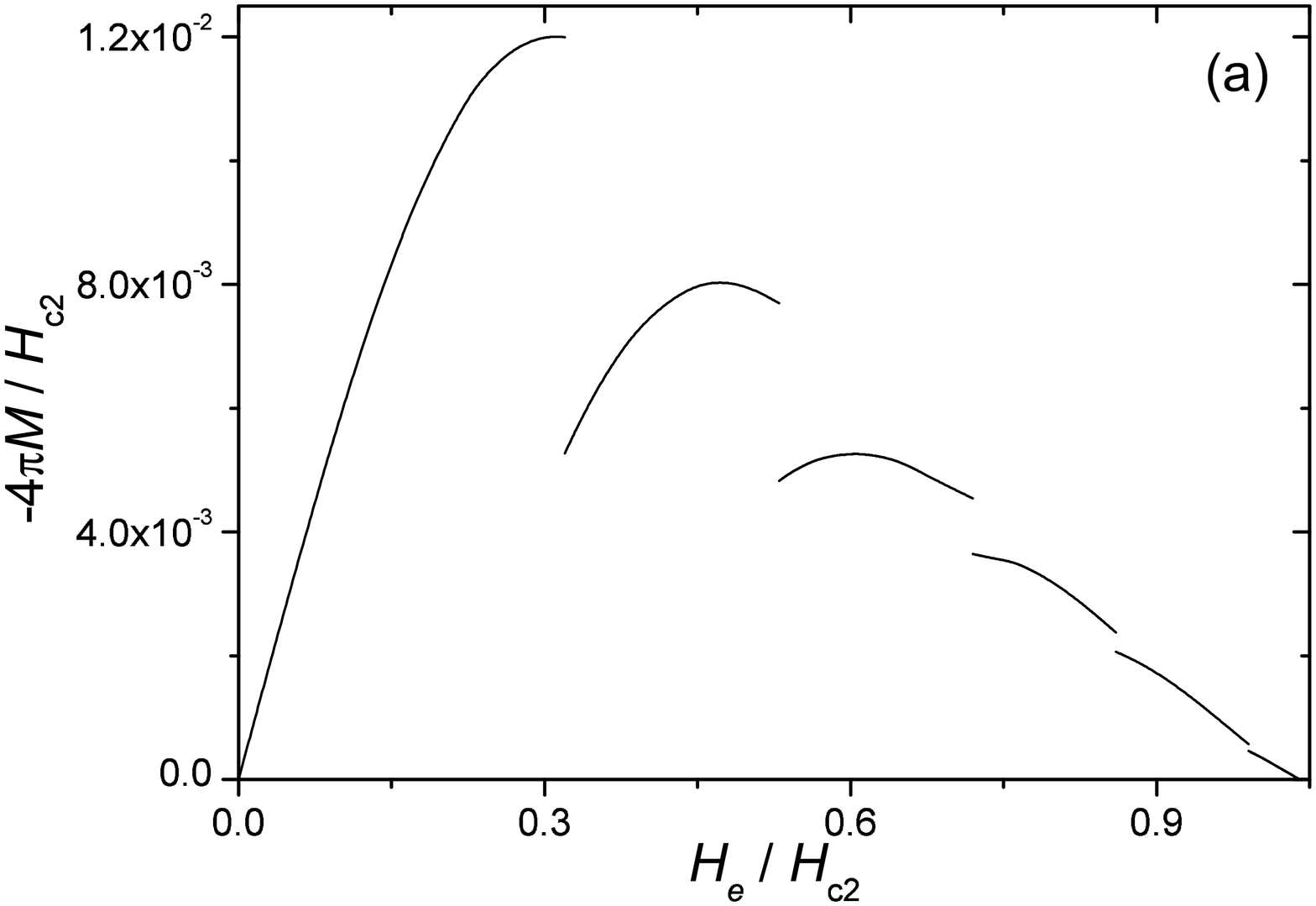}}
\text{Fig. 3(a)}
\epsfxsize= .85\hsize
\centerline{ \epsffile{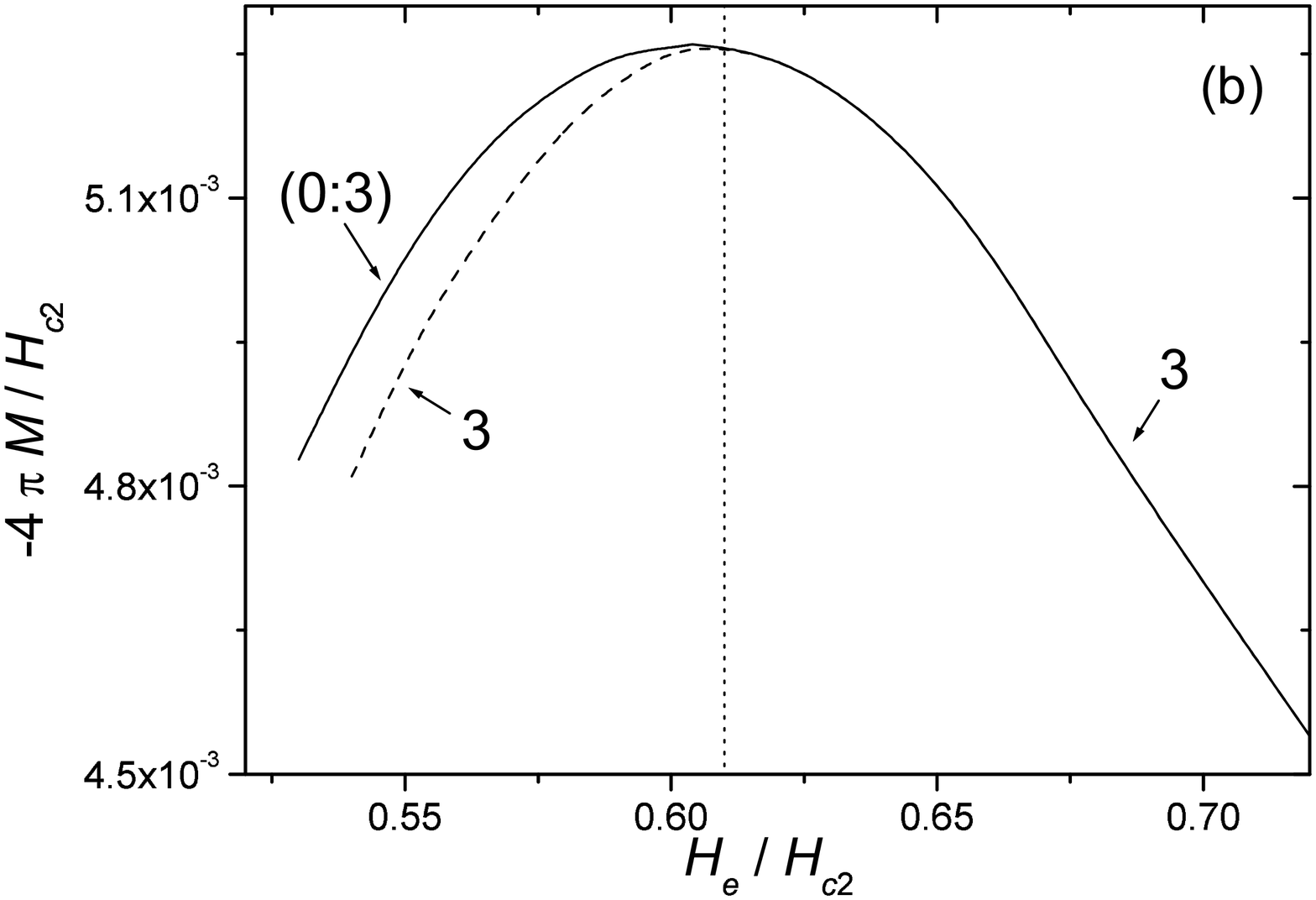}}
\text{Fig. 3(b)}
\end{figure}
\newpage
\begin{figure}[tbp]
\epsfxsize= 0.85\hsize
\centerline{ \epsffile{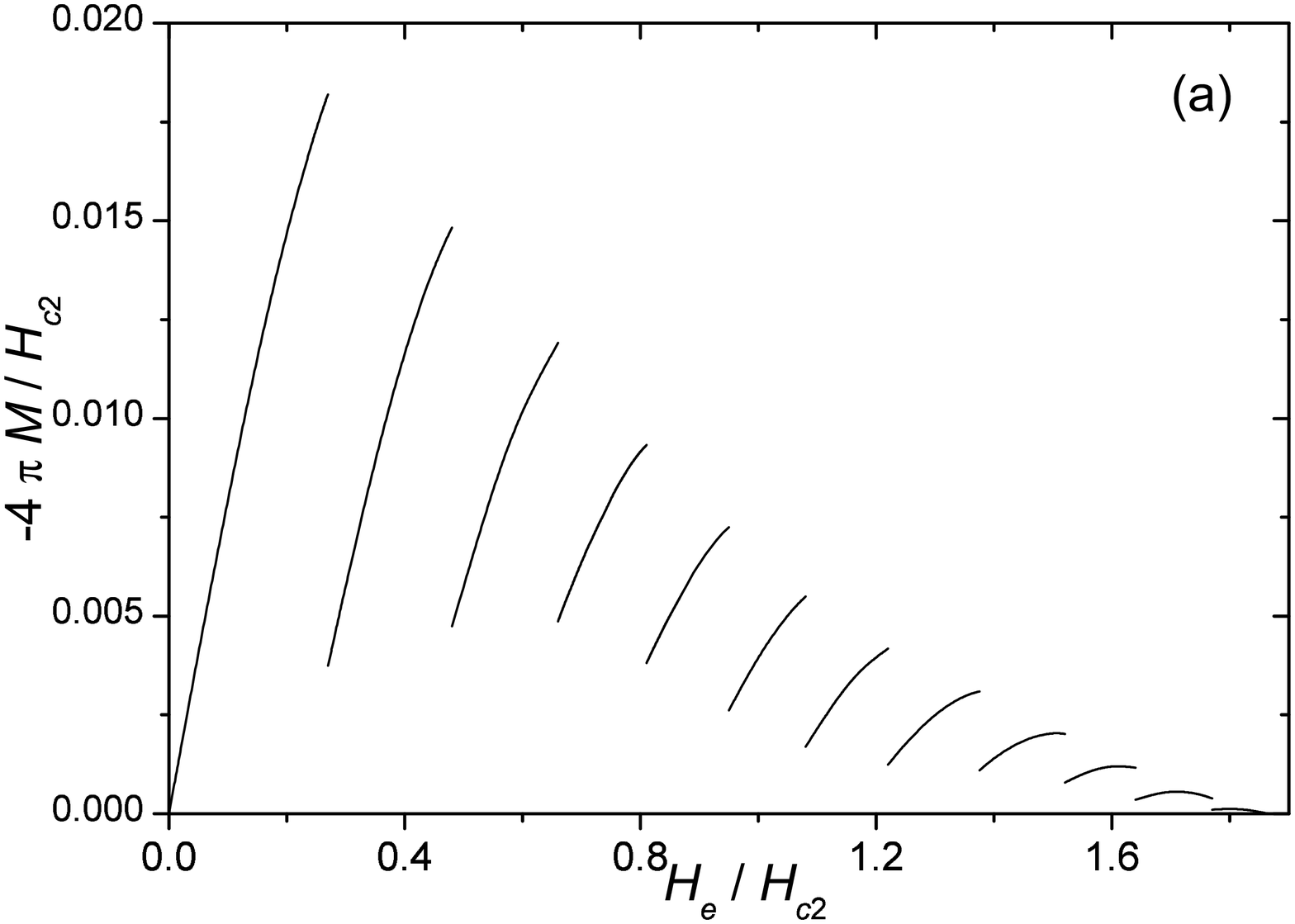}}
\text{Fig. 4(a)}
\epsfxsize= 0.85\hsize
\centerline{ \epsffile{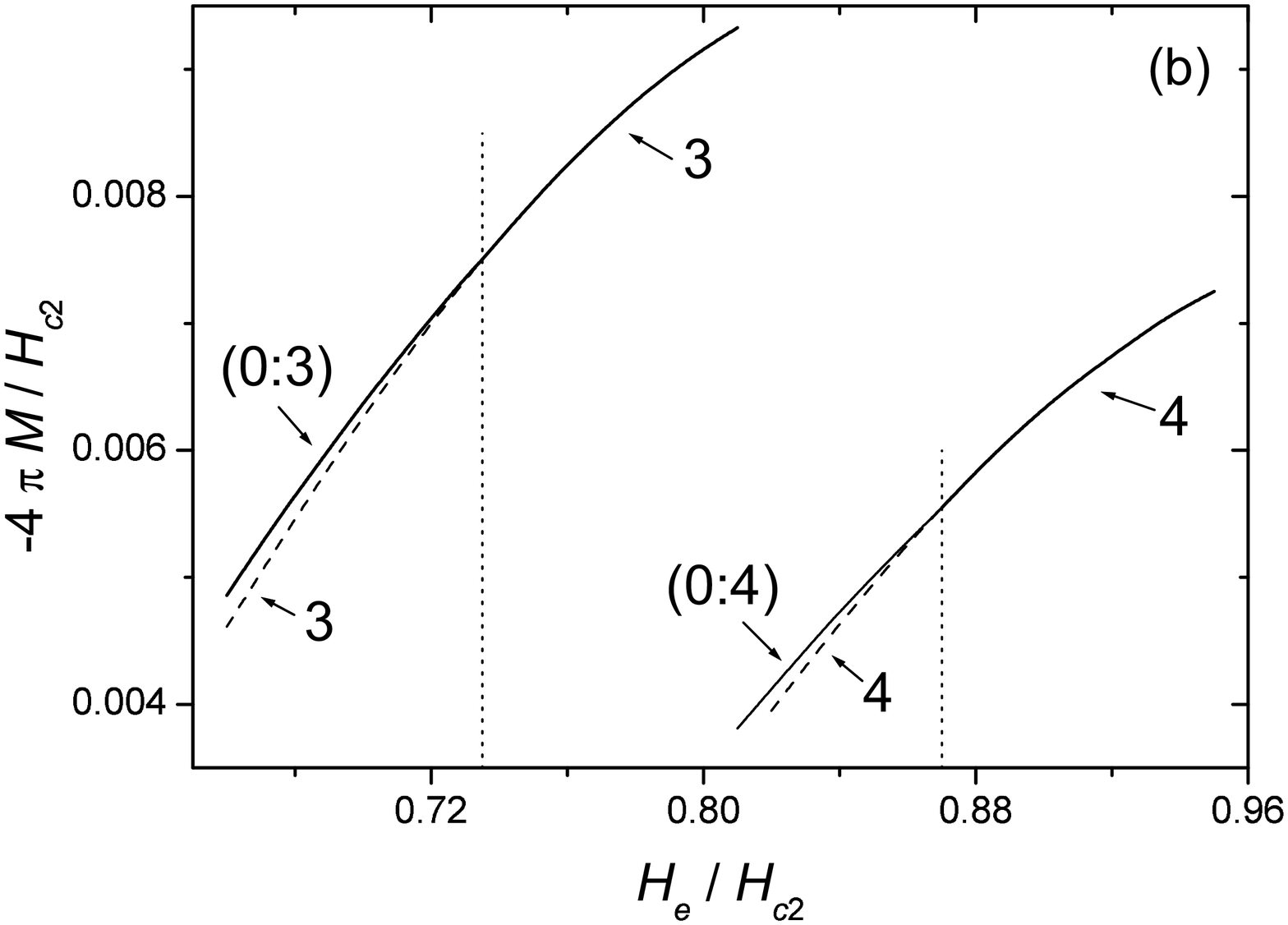}}
\text{Fig. 4(b)}
\end{figure}

\end{document}